\documentclass[journal,10pt,a4paper]{IEEEtran}
\usepackage[T1]{fontenc}
\usepackage{newtxtext}
\usepackage{amsmath,amssymb,bm}
\usepackage{newtxmath}
\usepackage{microtype}
\usepackage{graphicx,booktabs,array,tabularx,multirow}
\usepackage[nocompress]{cite}
\usepackage{float,algorithm,algpseudocode}
\usepackage[font=footnotesize,labelfont=normalfont,labelsep=period]{caption}
\usepackage{dblfloatfix}
\usepackage{placeins}

\usepackage[hidelinks]{hyperref}
\hypersetup{pdftitle={HARS: LDPC Bit-Flipping Decoding With Initial-Syndrome-Conditioned Parameter Mapping and Local-Reliability Weighting},pdfauthor={Tung Hsu Wu and Huang-Chang Lee},pdfsubject={HARS LDPC decoding and partially parallel FPGA architecture},pdfkeywords={HARS,Hybrid Adaptive Reliability-Aware and Syndrome-Aware,LDPC,bit-flipping,initial syndrome,local reliability,FPGA}}
\newcommand{\doi}[1]{\href{https://doi.org/#1}{\nolinkurl{#1}}}
\graphicspath{{figures/}}

\newcommand{\clip}[3]{\operatorname{clip}\!\left(#1,[#2,#3]\right)}
\newcommand{\rnd}{\operatorname{round}_{\mathrm{away}}}
\newcommand{\sat}{\operatorname{sat}}
\newcommand{\M}{\mathcal M}
\newcommand{\N}{\mathcal N}
\newcommand{\Fset}{\mathcal F}
\newcommand{\Ebn}{E_b/N_0}
\newcommand{\Sfmt}[2]{\mathrm{S}(#1,#2)}
\newcommand{\Ufmt}[2]{\mathrm{U}(#1,#2)}
\newcolumntype{L}[1]{>{\raggedright\arraybackslash}p{#1}}
\newcolumntype{Y}{>{\raggedright\arraybackslash}X}
\algrenewcommand\algorithmicindent{1.0em}
\title{HARS: LDPC Bit-Flipping Decoding With Initial-Syndrome-Conditioned Parameter Mapping and Local-Reliability Weighting}
\author{Tung Hsu Wu and Huang-Chang Lee\thanks{This work was supported by the National Science and Technology Council, Taiwan, under Grant 115-2813-C-019-028-E. (Corresponding author: Huang-Chang Lee.)}\thanks{Tung Hsu Wu and Huang-Chang Lee are with the Department of Electrical Engineering, National Taiwan Ocean University, Keelung, Taiwan (e-mail: 258258258258dd@gmail.com; hclee@mail.ntou.edu.tw).}}

\begin{document}
\maketitle
\begin{abstract}
HARS (Hybrid Adaptive Reliability-Aware and Syndrome-Aware) is a bit-flipping decoder that combines local channel reliability with parameter selection from the initial syndrome. A bounded check reliability modifies the parity-check weight, while the initial syndrome weight selects the base weight, reliability coefficient, threshold decay factor, and perturbation amplitude for each frame. Computing these quantities at initialization supports synchronous multi-bit updates with a simple iterative datapath. For rate-one-half PEGReg and WiMAX codes over the binary-input additive white Gaussian noise channel, HARS reduces both bit-error rate (BER) and mean iteration count relative to a calibrated SM-NGDBF baseline. The PEGReg BER at 3.25 dB is approximately one-sixth of the comparison value; the WiMAX BER at 2.75 dB is reduced by 59\%. A seven-fractional-bit FPGA implementation shares 132 variable-node lanes among four frame contexts. Precomputed check weights, encoded threshold states, and parallel perturbation generation support one group per clock during sustained decoding. On an XC7A200T at 60 MHz, for preloaded 128-frame batches with a ready receiver, measured coded throughputs are 49.7 Mb/s at 2.50 dB and 361.9 Mb/s at 4.00 dB, including initialization and output transfers.
\end{abstract}

\section{Introduction}
Bit-flipping (BF) decoding offers a compact alternative to soft-message decoding of low-density parity-check (LDPC) codes\cite{Gallager1962,MacKay1996,Kou2001}. Rather than exchanging belief-propagation or min-sum messages along every Tanner-graph edge\cite{Chen2002}, a BF decoder selects hard decisions for inversion from channel observations and parity-check outcomes. The central design problem is to extract enough reliability information for accurate node selection while keeping the update rule suitable for parallel hardware.

Weighted bit-flipping (WBF) and modified WBF distinguish check reliability through channel magnitudes\cite{Kou2001,Zhang2004}. Gradient descent bit flipping (GDBF) combines channel agreement and parity-check information in an inversion metric derived from a discrete objective\cite{Wadayama2010}. Noisy GDBF (NGDBF) adds perturbations to escape persistent decision patterns; its smoothed multi-bit form (SM-NGDBF) combines local threshold adaptation with a vote over terminal decisions\cite{Sundararajan2014}. These methods improve node selection without the soft-message storage required by belief propagation.

SM-NGDBF applies a common coefficient to all syndrome contributions. The associated checks, however, need not have the same channel reliability. WBF weights a check by its minimum adjacent channel magnitude\cite{Kou2001}, and new reliability-ratio weighted bit flipping (NRRWBF) relates this minimum to the magnitude at an individual variable node\cite{Aqil2021}. HARS uses a bounded check reliability $\rho_i$ and the weight $w+\alpha\rho_i$. The base weight preserves the parity-check contribution when the minimum magnitude is small; the additional term differentiates checks according to their channel neighborhoods.

The initial syndrome provides a second source of information. Its weight $S_0$ counts the parity constraints violated by the received hard decisions. In HARS, this count selects the base weight $w$, reliability coefficient $\alpha$, threshold decay factor $\lambda$, and perturbation amplitude $\eta$ for the frame. Local reliability then differentiates checks within that frame. Both quantities are available before decoding begins and remain fixed during the iterations.

Previous NGDBF improvements use inversion histories, syndrome information, and modified perturbations\cite{Dai2018,Deng2022,Li2022}. Dai \emph{et al.} adjust syndrome contributions using the preceding inversion metrics of neighboring nodes; their multi-bit method also sets a flip threshold from the current syndrome weight. Deng \emph{et al.} select a perturbation mean according to the time since the last flip and the sign of the channel correlation term. Li \emph{et al.} derive perturbations from channel samples and combine cyclic shifts with modified node-selection rules. HARS uses $S_0$ and $\rho_i$ at initialization, after which the decisions, syndromes, and node thresholds evolve through synchronous updates.

The hardware design exploits this separation between initialization and iteration. Precomputing check weights removes their multiplication from the iterative datapath, and encoding the rounded threshold recurrence replaces per-node threshold multiplication with table lookup. The resulting field-programmable gate array (FPGA) architecture shares 132 processing lanes among four frame contexts. Double-buffered decisions and syndromes preserve synchronous updates across the shared pipeline. Related BF implementations use stored inversion metrics, quasi-cyclic (QC) node shifting, or quantized perturbations\cite{Cui2020,Le2018,Li2025}; memory banking and frame interleaving also support parallel QC decoding\cite{KimPark2013,Mhaske2015,Hasani2022,Kumawat2015}.

The following sections develop the HARS update rule and its code-specific mappings, evaluate decoding performance on PEGReg and WiMAX codes, examine the two mechanisms through a PEGReg ablation, select the fixed-point precision, and present the architecture and FPGA results.

\section{System Model and Decoding Background}
\subsection{Channel, Decisions, and Syndromes}
Let $H\in\{0,1\}^{M\times N}$ be the parity-check matrix and let the codeword $\bm c$ satisfy $H\bm c=\bm0\pmod2$. The variable-node neighborhood of check node (CN) $i$ is $\N(i)=\{k:H_{ik}=1\}$. The adjacent check-node set of variable node (VN) $k$ is $\M(k)=\{i:H_{ik}=1\}$. Binary phase-shift keying (BPSK) maps $c_k$ to $1-2c_k$. The received sample is
\begin{equation}
 y_k=1-2c_k+\nu_k,\qquad \nu_k\sim\mathcal N(0,v_n),
 \label{eq:channel}
\end{equation}
where the channel variance for unit symbol energy and code rate $R$ is
\begin{equation}
 v_n=\frac{1}{2R\,10^{(\Ebn)_{\mathrm{dB}}/10}}.
 \label{eq:variance}
\end{equation}
The experiments transmit the all-zero codeword. Bit and frame errors are counted by comparing the output with that transmitted word, separately from the parity-based stopping test.

Channel samples are clipped as
\begin{equation}
 \bar y_k=\clip{y_k}{-Y_{\max}}{Y_{\max}},
 \label{eq:clip}
\end{equation}
with $\operatorname{clip}(u,[a,b])=\min(\max(u,a),b)$. The initial hard decision is $\hat c_k^{(0)}=\mathbf1[\bar y_k<0]$, and the bipolar decision is $x_k^{(t)}=1-2\hat c_k^{(t)}$. Binary and bipolar syndromes are
\begin{align}
 z_i^{(t)}&=\bigoplus_{j\in\N(i)}\hat c_j^{(t)},\\
 s_i^{(t)}&=1-2z_i^{(t)}.
 \label{eq:syndrome}
\end{align}
A satisfied check has $s_i=+1$, and an unsatisfied check has $s_i=-1$. When the binary syndrome is zero, the current decision satisfies the code constraints and decoding stops.

\subsection{From GDBF to Smoothed Multi-Bit NGDBF}
GDBF combines the channel observations and parity constraints in the objective\cite{Wadayama2010}
\begin{equation}
 f(\bm x)=\sum_{k=1}^{N}x_ky_k+
 \sum_{i=1}^{M}\prod_{j\in\N(i)}x_j.
 \label{eq:objective}
\end{equation}
The first sum measures channel agreement. Each term in the second sum is $+1$ for a satisfied check and $-1$ otherwise. Flipping only bit $k$ reverses its channel term and all adjacent check terms, giving the inversion metric
\begin{equation}
 E_k^{\mathrm{GDBF}}=x_ky_k+\sum_{i\in\M(k)}s_i.
 \label{eq:gdbf}
\end{equation}
For a single-bit flip yielding $\bm x'$, $f(\bm x')-f(\bm x)=-2E_k^{\mathrm{GDBF}}$; a negative inversion metric therefore increases the objective.

For fixed received samples, returning to the same decision state also returns to the same GDBF metric. NGDBF adds fresh perturbations so that nodes near a flip boundary can be selected differently\cite{Sundararajan2014}:
\begin{equation}
 E_k^{\mathrm{NGDBF}}=x_k\bar y_k+w\sum_{i\in\M(k)}s_i
 +\eta\sqrt{v_n}\,q_k,
 \label{eq:ngdbf}
\end{equation}
where $q_k\sim\mathcal N(0,1)$, $w$ is the check weight, and $\eta$ is the noise-scale coefficient of the original method. 

Single-bit NGDBF selects the node with the minimum metric. Multi-bit NGDBF flips all nodes satisfying $E_k<\theta_k$, with a local threshold $\theta_k$ that evolves with the node's update history. One iteration evaluates all nodes from the same pre-update decisions and syndromes, then commits the selected flips synchronously.

SM-NGDBF forms its output from a vote over the terminal decision history when the iteration limit is reached\cite{Sundararajan2014}. A zero syndrome instead terminates decoding immediately with the current decisions.

\section{The HARS Decoder}
\subsection{Local-Reliability Weighting}
A parity-check result depends on all adjacent variable-node decisions. WBF uses their minimum channel magnitude for check weighting\cite{Kou2001}, and NRRWBF includes this minimum in a reliability ratio\cite{Aqil2021}. HARS defines a bounded check-level measure as
\begin{equation}
 \rho_i=\frac{\min\!\left(\min_{j\in\N(i)}|\bar y_j|,\rho_{\max}\right)}{\rho_{\max}},
 \qquad 0\leq\rho_i\leq1.
 \label{eq:rho}
\end{equation}
The neighborhood minimum reflects the weakest channel observation associated with the check. Clipping at $\rho_{\max}$ bounds the reliability correction and gives a fixed range for its hardware representation.

The effective check-node weight is defined by
\begin{equation}
 W_i\triangleq w(S_0)+\alpha(S_0)\rho_i,
 \label{eq:effective_weight}
\end{equation}
where $w$ is the base parity-check weight and $S_0$ is defined below. For $\alpha\geq0$, $W_i\in[w,w+\alpha]$: weak channel support leaves the base contribution intact, while stronger support increases the magnitude of the check contribution.

The syndrome sign determines the effect of this weighting. A reliable satisfied check increases the inversion metric and favors retaining the decision; a reliable unsatisfied check favors a flip. Equal counts of unsatisfied checks can therefore lead to different node selections when their channel reliabilities differ.

\subsection{Initial-Syndrome-Conditioned Parameter Mapping}
Local reliability distinguishes checks within a frame. The initial syndrome determines the frame-level update scale. Define
\begin{equation}
 S_0=\sum_{i=1}^{M}z_i^{(0)}.
 \label{eq:s0}
\end{equation}
For each $g\in\{w,\alpha,\lambda,\eta\}$, HARS uses a clipped linear mapping
\begin{equation}
 g(S_0)=\clip{a_gS_0+b_g}{g_{\min}}{g_{\max}}.
 \label{eq:mapping}
\end{equation}
Here $w$ is the base parity-check weight, $\alpha$ is the local-reliability coefficient, $\lambda$ is the threshold decay factor, and $\eta$ is the perturbation amplitude. All four are evaluated at initialization and shared by every node and update in that frame.

\begin{figure*}[t]
 \centering
 \includegraphics[width=0.75\textwidth]{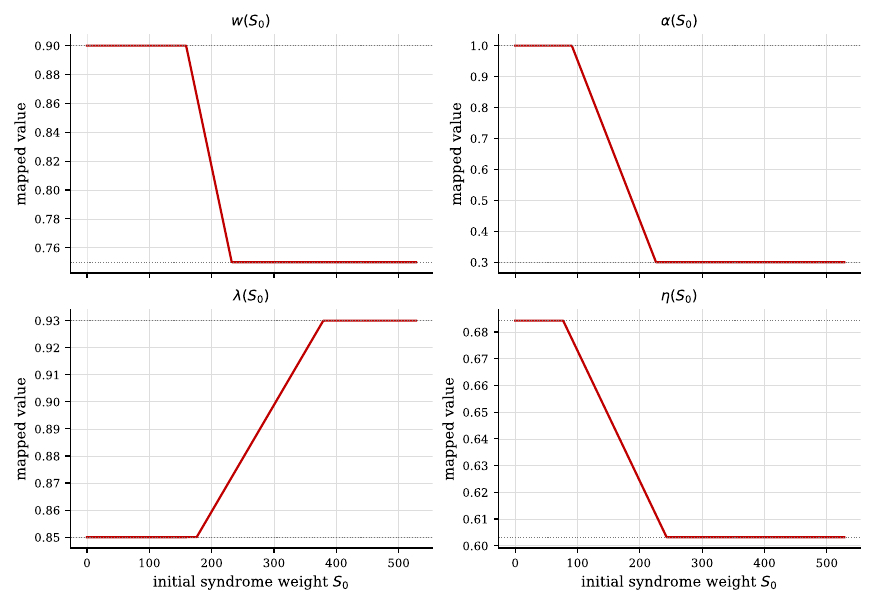}
 \caption{Initial-syndrome-conditioned mappings for WiMAX HARS. Solid curves are the mapped values; dashed lines mark the clipping limits. All four parameters are evaluated from the same frame's $S_0$.}
 \label{fig:mapping}
\end{figure*}

Fig.~\ref{fig:mapping} shows the jointly calibrated WiMAX mappings. Increasing $S_0$ reduces $w$, $\alpha$, and $\eta$, while increasing $\lambda$ slows the movement of negative thresholds toward zero. The mappings thus adjust both the inversion metric and threshold evolution to the initial syndrome.

\begin{table}[!htbp]
\centering
\caption{HARS Parameter Mappings and Common Controls}
\label{tab:mapping}
\footnotesize
\setlength{\tabcolsep}{2.5pt}
\renewcommand{\arraystretch}{1.1}
\begin{tabular*}{\columnwidth}{@{\extracolsep{\fill}}llrrrr@{}}
\toprule
Code & Parameter & $a_g$ & $b_g$ & $g_{\min}$ & $g_{\max}$ \\
\midrule
PEGReg & $w$ & $-2.030\times10^{-3}$ & 1.013 & 0.400 & 0.950 \\
 & $\alpha$ & $1.330\times10^{-3}$ & 0.5250 & 0 & 1.20 \\
 & $\lambda$ & $-7.500\times10^{-4}$ & 1.023 & 0.800 & 0.990 \\
 & $\eta$ & $-2.320\times10^{-3}$ & 0.9422 & 0.390 & 0.759 \\
\midrule
WiMAX & $w$ & $-2.051\times10^{-3}$ & 1.226 & 0.750 & 0.900 \\
 & $\alpha$ & $-5.189\times10^{-3}$ & 1.472 & 0.300 & 1.00 \\
 & $\lambda$ & $3.937\times10^{-4}$ & 0.7807 & 0.850 & 0.930 \\
 & $\eta$ & $-4.882\times10^{-4}$ & 0.7218 & 0.603 & 0.684 \\
\bottomrule
\end{tabular*}
\par\smallskip
\begin{minipage}{\columnwidth}\footnotesize
Both codes use $T_{\max}=300$, $Y_{\max}=2.5$, $\rho_{\max}=1.5$, and $L=64$. The thresholds $(\theta_0,\theta_{\mathrm{reset}})$ are $(-0.900,-0.385)$ for PEGReg and approximately $(-0.998,-0.542)$ for WiMAX. Displayed real coefficients are rounded. The accompanying parameter file lists the full-precision WiMAX coefficients and integer constants for F7 (seven fractional bits).
\end{minipage}
\end{table}

\begin{figure*}[t]
 \centering
 \includegraphics[width=0.95\textwidth]{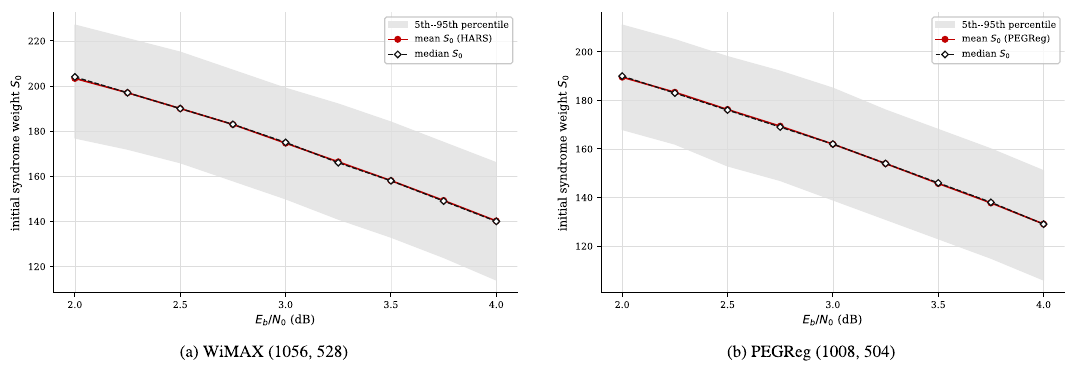}
 \caption{Initial syndrome weights for WiMAX and PEGReg. Solid and dashed curves denote the mean and median, respectively. Shaded bands show the 5th--95th percentiles across received frames.}
 \label{fig:s0}
\end{figure*}

As shown in Fig.~\ref{fig:s0}, the mean and median $S_0$ decrease with increasing $\Ebn$, but the received frames span a range of initial syndrome weights at each operating point. Evaluating the mappings for each frame uses this variation directly, while $\rho_i$ retains the check-level channel information.

The coefficients in Table~\ref{tab:mapping} are calibrated separately for each code. In particular, $\alpha$ increases with $S_0$ for PEGReg and decreases for WiMAX.

\subsection{Parameter Formation and Code-Specific Settings}
PEGReg calibration starts from the SM-NGDBF comparison setting and retains its channel clipping, iteration limit, threshold initialization and reset, and smoothing controls. Successive coordinate-wise grid searches jointly calibrate $w$, $\alpha$, $\lambda$, and $\eta$: each search refines one parameter interval using BER while holding the others at their current values.

At each operating point, averages of the better-performing parameter candidates and their corresponding $S_0$ data define representative settings for linear fitting. Clipping the fitted outputs gives (\ref{eq:mapping}). The fitted coefficients remain fixed during BER evaluation; decoding uses only the received frame's $S_0$ to select the four parameters.

For WiMAX, HARS and SM-NGDBF are calibrated separately using the WiMAX matrix and initial-syndrome distribution. The HARS reliability measure, mapping form, and synchronous update rule are unchanged. Its QC structure is also used to organize the hardware interconnect.

\subsection{Inversion Metric and Synchronous Flipping}
HARS combines local weighting and frame-level control as
\begin{equation}
\begin{split}
 E_k^{\mathrm{HARS}}={}&x_k\bar y_k+
 \sum_{i\in\M(k)}W_is_i\\
 &+\eta(S_0)q_k.
\end{split}
\label{eq:hars}
\end{equation}
A fresh standard-normal perturbation $q_k$ is drawn for each node and iteration. The frame parameters follow (\ref{eq:mapping}), with $\lambda$ controlling the threshold recurrence. The fixed-point implementation uses the discrete perturbations evaluated in Section~\ref{sec:precision_selection}.

With the dependence on $S_0$ left implicit, the check contribution can be separated into base and reliability terms:
\begin{equation}
\begin{split}
 E_k^{\mathrm{HARS}}={}&x_k\bar y_k+w\sum_{i\in\M(k)}s_i\\
 &+\alpha\sum_{i\in\M(k)}\rho_i s_i+\eta q_k.
\end{split}
\label{eq:hars_expanded}
\end{equation}
Both check terms use the same syndrome sign, so local reliability changes the strength of the parity information without changing its direction.

Let $\hat{\bm c}^{(t-1)}$ and $\bm\theta^{(t-1)}$ be the state before update $t$. The value $E_k^{(t)}$ is obtained from (\ref{eq:hars}) using $x_k^{(t-1)}$, $s_i^{(t-1)}$, and the new sample $q_k^{(t)}$. Define
\begin{equation}
 \Fset^{(t)}=\{k:E_k^{(t)}<\theta_k^{(t-1)}\}.
 \label{eq:flipset}
\end{equation}
After all comparisons, decisions are committed synchronously:
\begin{equation}
 \hat c_k^{(t)}=\hat c_k^{(t-1)}\oplus\mathbf1[k\in\Fset^{(t)}].
 \label{eq:commit}
\end{equation}
Equality retains the prior decision. Forming the complete flip set before recomputing the syndrome preserves a common pre-update state for all nodes.

\subsection{Threshold Evolution and Terminal Smoothing}
Thresholds are initialized to $\theta_k^{(0)}=\theta_0$ and updated by
\begin{equation}
 \theta_k^{(t)}=
 \begin{cases}
  \theta_{\mathrm{reset}},&k\in\Fset^{(t)},\\
  \lambda(S_0)\theta_k^{(t-1)},&k\notin\Fset^{(t)}.
 \end{cases}
 \label{eq:threshold}
\end{equation}
For a retained node, $0<\lambda<1$ moves the negative threshold toward zero and gradually relaxes the flip condition. A flipped node returns to $\theta_{\mathrm{reset}}$.

A zero initial or post-update syndrome returns the current decisions immediately. A frame reaching $T_{\max}$ with a nonzero syndrome uses the final $L=64$ committed decisions. Its bit-wise vote is
\begin{equation}
 v_k=\sum_{t=T_{\max}-L+1}^{T_{\max}}(1-2\hat c_k^{(t)}).
 \label{eq:votes}
\end{equation}
The output is zero for $v_k>0$ and one for $v_k<0$; a tie retains the decision from update $T_{\max}$. Accumulation starts at update $T_{\max}-L+1$ and uses the post-commit decisions. The final syndrome is recomputed after resolving all votes. Algorithm~\ref{alg:hars} specifies the complete procedure.

\begin{algorithm}[t]
\caption{Synchronous Multi-Bit HARS Decoding}
\label{alg:hars}
\fontsize{8.1}{10.0}\selectfont
\begin{algorithmic}[1]
\Statex \textbf{Input:} $\bm y$, $H$, mapping coefficients and clipping limits;
\Statex \hspace{1em}$Y_{\max}$, $\rho_{\max}$, $\theta_0$, $\theta_{\mathrm{reset}}$, $T_{\max}$, $L$
\Statex \textbf{Output:} decisions, parity status, updates, syndrome weight
\State $\bar y_k\gets\clip{y_k}{-Y_{\max}}{Y_{\max}}$, $k=1,\ldots,N$
\State $\hat c_k^{(0)}\gets\mathbf1[\bar y_k<0]$
\State $\bm z^{(0)}\gets H\hat{\bm c}^{(0)}\pmod2$; $S_0\gets\sum_i z_i^{(0)}$
\If{$S_0=0$}
 \State \textbf{return} $\hat{\bm c}^{(0)}$, parity satisfied, $0$, $0$
\EndIf
\For{$g\in\{w,\alpha,\lambda,\eta\}$}
 \State $g\gets\clip{a_gS_0+b_g}{g_{\min}}{g_{\max}}$
\EndFor
\State $\rho_i\gets\min(\min_{j\in\N(i)}|\bar y_j|,\rho_{\max})/\rho_{\max}$, all $i$
\State $\theta_k^{(0)}\gets\theta_0$; $v_k\gets0$, $k=1,\ldots,N$
\For{$t=1,\ldots,T_{\max}$}
 \State $x_k^{(t-1)}\gets1-2\hat c_k^{(t-1)}$; $s_i^{(t-1)}\gets1-2z_i^{(t-1)}$
 \For{$k=1,\ldots,N$}
  \State Generate a fresh $q_k^{(t)}\sim\mathcal N(0,1)$
  \State Evaluate (\ref{eq:hars}) with $x_k^{(t-1)}$, $s_i^{(t-1)}$, and $q_k^{(t)}$
 \EndFor
 \State $\Fset^{(t)}\gets\{k:E_k^{(t)}<\theta_k^{(t-1)}\}$
 \State Commit $\hat c_k^{(t)}\gets\hat c_k^{(t-1)}\oplus\mathbf1[k\in\Fset^{(t)}]$, all $k$
 \For{$k=1,\ldots,N$}
  \If{$k\in\Fset^{(t)}$}
   \State $\theta_k^{(t)}\gets\theta_{\mathrm{reset}}$
  \Else
   \State $\theta_k^{(t)}\gets\lambda\theta_k^{(t-1)}$
  \EndIf
 \EndFor
 \State $\bm z^{(t)}\gets H\hat{\bm c}^{(t)}\pmod2$
 \If{$t\geq T_{\max}-L+1$}
  \State $v_k\gets v_k+(1-2\hat c_k^{(t)})$, $k=1,\ldots,N$
 \EndIf
 \If{$\sum_i z_i^{(t)}=0$}
  \State \textbf{return} $\hat{\bm c}^{(t)}$, parity satisfied, $t$, $0$
 \EndIf
\EndFor
\For{$k=1,\ldots,N$}
 \State $\tilde c_k\gets0$ if $v_k>0$; $\tilde c_k\gets1$ if $v_k<0$
 \State If $v_k=0$, set $\tilde c_k\gets\hat c_k^{(T_{\max})}$
\EndFor
\State $\bm z^{\mathrm{final}}\gets H\tilde{\bm c}\pmod2$
\State \textbf{return} $\tilde{\bm c}$, $\sum_i z_i^{\mathrm{final}}=0$, $T_{\max}$, $\sum_i z_i^{\mathrm{final}}$
\end{algorithmic}
\end{algorithm}

\subsection{State and Computational Structure}
HARS stores four frame parameters and $M$ check reliabilities, all fixed after initialization. The evolving state comprises $N$ decisions, $N$ thresholds, $M$ syndromes, and one smoothing counter per bit.

Let $E_H=\sum_k|\M(k)|$ be the number of Tanner-graph edges. Initialization scans the edges for syndrome and minimum calculations and evaluates four multiply-add-and-clip mappings. Each update scans the neighborhoods for weighted checks and syndrome recomputation, together with $N$ metric, threshold, and decision operations. At fixed word length, the operation count is $O(E_H+N+M)$ per iteration and $O((I+1)(E_H+N+M))$ for a frame that executes $I$ iterations, including initialization. Local weighting enters neighborhood accumulation, and flipping is decided by a per-node threshold comparison, allowing the arithmetic to be scheduled on a shared datapath.

\section{Decoding Performance}
\subsection{Experimental Setup}
The experiments use the rate-one-half PEGReg $(1008,504)$ and WiMAX $(1056,528)$ codes. PEGReg connects the evaluation to the NGDBF benchmark\cite{Sundararajan2014}. WiMAX uses a $12\times24$ base matrix with circulant size $Z=44$\cite{IEEE80216}. Floating-point tests cover $\Ebn=2.0$--$4.0$~dB in 0.25-dB steps.

At each operating point, the methods use a common channel-frame sequence, with channel noise generated separately from decoder perturbations. Error-count and sample limits determine the prefix decoded by each method, with extended sampling at high-SNR HARS points. BER is the observed bit-error count divided by the decoded-bit count. Mean iterations include every decoded frame, assigning zero updates to an initially zero syndrome and $T_{\max}$ to a frame reaching the limit.

The SM-NGDBF comparison follows the perturbation and smoothing structure of \cite{Sundararajan2014}, with the code-specific settings in Table~\ref{tab:sm}. Its threshold is reset after a flip, whereas the original rule retains the threshold on a flip. HARS and this comparison variant share $T_{\max}=300$, $Y_{\max}=2.5$, and a 64-decision smoothing window.

\begin{table}[!htbp]
\centering
\caption{SM-NGDBF Comparison Settings}
\label{tab:sm}
\small
\renewcommand{\arraystretch}{1.1}
\begin{tabular*}{\columnwidth}{@{\extracolsep{\fill}}lrr@{}}
\toprule
Parameter & PEGReg & WiMAX \\
\midrule
$w$ & 0.750 & 0.994 \\
$\eta$ & 0.950 & 0.876 \\
$\theta_0$ & $-0.900$ & $-1.83$ \\
$\theta_{\mathrm{reset}}$ & $-0.385$ & $-0.970$ \\
$\lambda$ & Piecewise & 0.788 \\
\bottomrule
\end{tabular*}
\par\smallskip
\begin{minipage}{\columnwidth}\footnotesize
PEGReg uses $\lambda=0.990$, 0.970, and 0.940 over $\Ebn\leq3.25$, $3.25<\Ebn\leq3.75$, and $3.75<\Ebn\leq4.0$~dB, respectively. Both codes use $T_{\max}=300$ and $L=64$. Perturbations follow (\ref{eq:ngdbf}).
\end{minipage}
\end{table}

PEGReg also includes single-bit adjustment-factor NGDBF (S-A-NGDBF)\cite{Dai2018} and normalized min-sum (NMS)\cite{Chen2002}, with iteration limits of 100 and 10. S-A-NGDBF uses $\lambda_1=-0.3$, $\lambda_2=1$, and adjustment-factor threshold $\theta_1=0$ from \cite{Dai2018}; the NMS comparison in this study uses a scaling factor of approximately 0.844. WiMAX additionally includes single-bit NGDBF (S-NGDBF) with $T_{\max}=300$, check weight 0.732, and noise scale 0.678. The iteration limits are shown in the legends.

\subsection{PEGReg Results}
\begin{figure}[!htbp]
 \centering
 \includegraphics[width=\columnwidth]{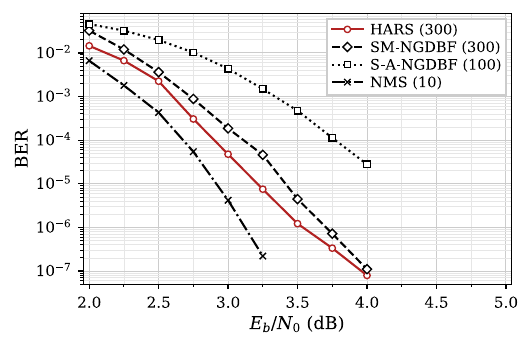}
 \caption{BER for PEGReg $(1008,504)$. Simulated points span 2.0--4.0~dB; parentheses specify the iteration limit. HARS and SM-NGDBF use synchronous multi-bit updates.}
 \label{fig:peg}
\end{figure}

HARS has the lowest BER among the three BF methods across the PEGReg sweep in Fig.~\ref{fig:peg}. At 3.25~dB, its BER is about one-sixth of the SM-NGDBF value. NMS provides the soft-message reference.

This BER reduction is accompanied by fewer iterations: at 3~dB, HARS reduces the mean iteration count by approximately 41\% relative to SM-NGDBF. Both decoders use synchronous multi-bit updates and the same iteration limit.

\subsection{WiMAX Results}
\begin{figure}[!htbp]
 \centering
 \includegraphics[width=\columnwidth]{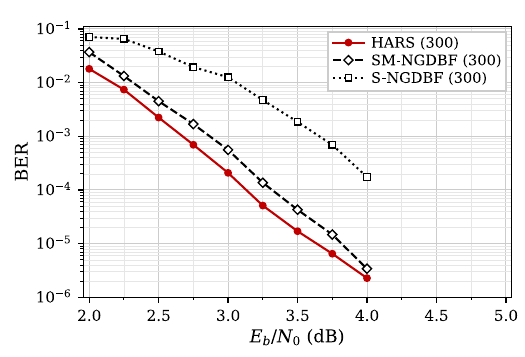}
 \caption{BER for WiMAX $(1056,528)$. Simulated points span 2.0--4.0~dB. All three methods have an iteration limit of 300.}
 \label{fig:wimax}
\end{figure}

The WiMAX results in Fig.~\ref{fig:wimax} show the same advantage over SM-NGDBF and S-NGDBF across the tested SNR range. At 2.75~dB, HARS reduces BER by approximately 59\% relative to SM-NGDBF.

\begin{figure}[!htbp]
 \centering
 \includegraphics[width=\columnwidth]{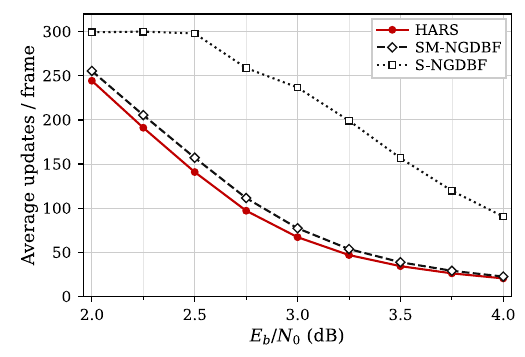}
 \caption{Mean iterations for WiMAX, including all tested frames. HARS and SM-NGDBF use synchronous multi-bit updates; S-NGDBF flips one node per update.}
 \label{fig:updates}
\end{figure}

HARS also requires fewer mean iterations than SM-NGDBF at all nine WiMAX points (Fig.~\ref{fig:updates}), with a reduction of approximately 13\% at 2.75~dB. The two multi-bit curves converge as SNR increases. The S-NGDBF curve uses single-bit updates.

\subsection{Ablation of Local Reliability and Frame Adaptation}
\begin{figure*}[t]
 \centering
 \includegraphics[width=0.76\textwidth]{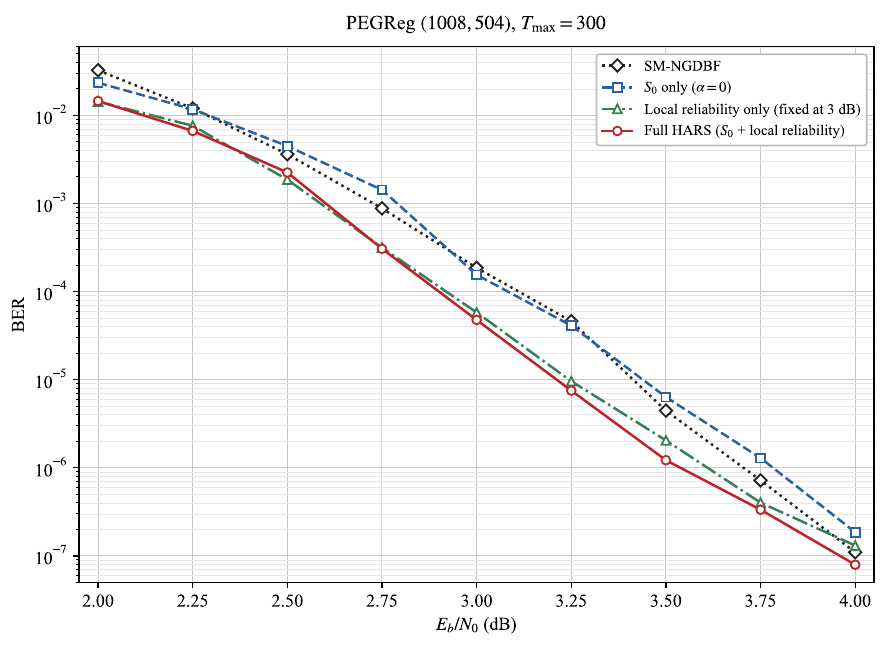}
 \caption{PEGReg BER ablation. The $S_0$-only configuration sets $\alpha=0$. The local-reliability-only configuration uses the same four constants, obtained from the mean $S_0$ at 3~dB, throughout the sweep.}
 \label{fig:ablation}
\end{figure*}
Figure~\ref{fig:ablation} compares full HARS on PEGReg, $S_0$ adaptation alone, and local reliability alone, together with the preceding SM-NGDBF baseline. The three HARS configurations share the matrix, clipping limits, initial and reset thresholds, 300-iteration limit, and 64-decision smoothing window.

The $S_0$-only configuration sets $\alpha=0$ and retains the full HARS mappings for $w(S_0)$, $\lambda(S_0)$, and $\eta(S_0)$. The local-reliability-only configuration retains $w+\alpha\rho_i$ and fixes all four parameters at the center of the tested range. The mean initial syndrome from 10,000 channel frames at 3~dB is $\bar S_{0,3}\approx162$. Evaluating the mappings gives
\begin{equation}
\begin{split}
 (w_{\mathrm{ref}},\alpha_{\mathrm{ref}},\lambda_{\mathrm{ref}},\eta_{\mathrm{ref}})
 \approx(0.684,0.740,0.902,0.566).
\end{split}
\label{eq:ablationref}
\end{equation}
The constants remain fixed over the sweep, while $\rho_i$ is evaluated from each received frame. Both ablated configurations retain the other HARS settings without retuning and use common frame-indexed channel samples and base perturbation sequences. Each run accumulates at least 250 bit errors and an operating-region-dependent target of 15--30 erroneous frames, counting all completed batches.

Local reliability accounts for the main BER reduction in Fig.~\ref{fig:ablation}. Frame-dependent parameter selection provides a further reduction of approximately 22\% at 3.25~dB, and full HARS remains below the fixed-parameter curve from 3.25 to 4~dB.

At 3~dB, the 5th--95th percentiles of $S_0$ span 139--185 around the mean used for the fixed setting. Frame adaptation adjusts the parameters across this range before local threshold comparisons determine the flip set.

A separate fixed-length comparison at 3~dB uses 100,000 common channel frames and base perturbation sequences. With the same unretuned configurations, frame adaptation reduces empirical BER by 4.6\% relative to local reliability with fixed parameters.

\subsection{Fixed-Point Precision Selection}
\label{sec:precision_selection}
Let $F$ denote the number of fractional bits in the main datapath. F3--F7 use $F=3,\ldots,7$, with a step size of $2^{-F}$ and common real-valued clipping limits. The four mapping parameters retain ten fractional bits.

Linear-feedback shift registers (LFSRs) generate five-level perturbations $\{-2,-1,0,1,2\}$, scaled by the frame-dependent amplitude; the generator is detailed in Section~\ref{sec:perturbations}.

\begin{figure*}[!t]
 \centering
 \includegraphics[width=0.72\textwidth]{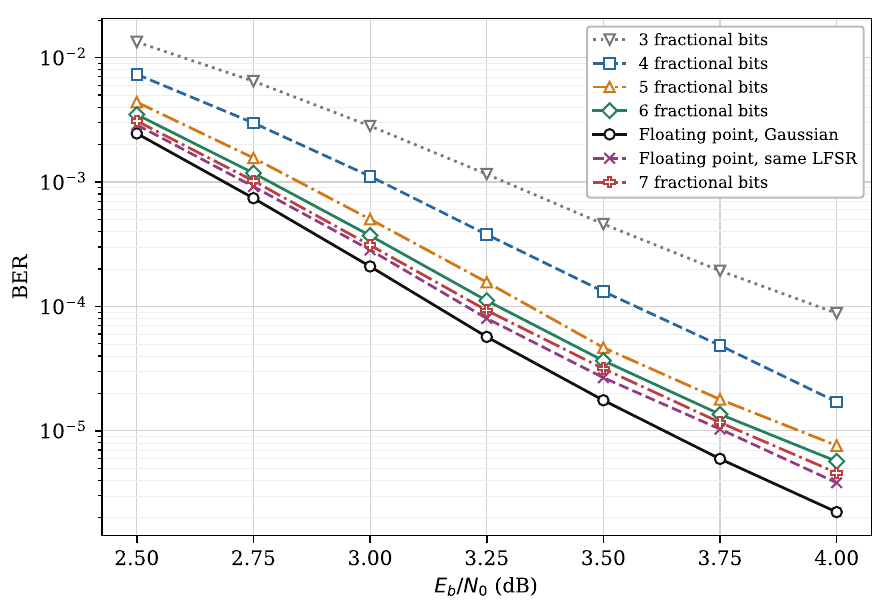}
 \caption{WiMAX precision and perturbation comparison. F3--F7 use three to seven fractional bits. One floating-point reference uses Gaussian perturbations; the other uses the same five-level LFSR sequence as the fixed-point configurations.}
 \label{fig:precision}
\end{figure*}

Figure~\ref{fig:precision} compares F3--F7 with two floating-point references. The Gaussian reference evaluates the original perturbation model; the same-LFSR reference isolates finite-precision effects by sharing the discrete sequence, seeds, and sample order with F3--F7. At 4.00 dB, F7 achieves approximately $4.60\times10^{-6}$ BER, about one-nineteenth of F3 and 19\% below F6. F7 is therefore selected for the architecture developed next.

The precision comparison uses common channel observations: 320,000 frames per point from 2.50 to 3.75 dB and 1,280,000 at 4.00 dB. The accompanying F7 dataset contains 2,070,000 frames across seven operating points; its per-point counts are supplied in the supplement.

\section{Partially Parallel HARS Architecture}
\setlength{\abovedisplayskip}{6pt plus 2pt minus 2pt}
\setlength{\belowdisplayskip}{6pt plus 2pt minus 2pt}
Initialization prepares the quantities that remain fixed within a frame; the iterative datapath then updates its decisions and syndromes. This separation moves check-weight products out of the decoding loop. Four frame contexts share a 132-lane datapath, allowing work from another frame to proceed while an iteration finishes writing back.

\subsection{Architecture Overview and Streaming Initialization}
\begin{figure*}[!t]
 \centering
 \includegraphics[width=0.92\textwidth]{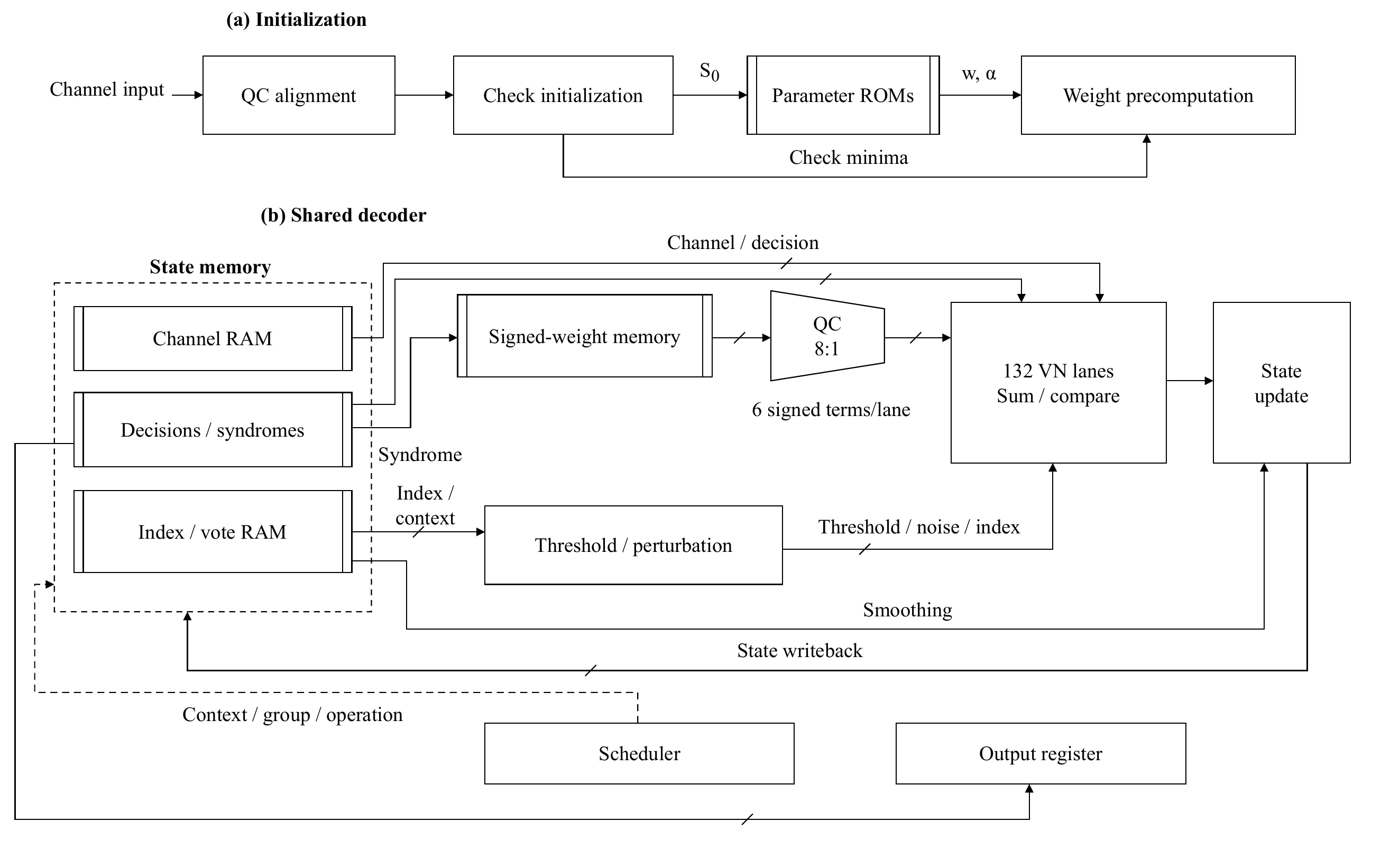}
 \caption{Shared decoder dataflow. (a) Initialization stores the channel samples and initial syndrome in the selected context and the precomputed signed weights in weight memory. (b) The same context supplies channel values, decisions, thresholds, and perturbations to 132 VN lanes. The scheduler selects the context, group, and operation; state updates return to memory. Each lane receives six signed check contributions, with unused slots set to zero. Smoothing bypasses VN arithmetic, and completed decisions use the shared state-read path to reach the output register.}
 \label{fig:core}
\end{figure*}

Figure~\ref{fig:core} shows the two paths. The front end receives channel samples, forms the initial syndrome and check minima, and precomputes signed check weights. During decoding, state and weight memories feed the VN lanes, which add the check contributions, channel term, and perturbation before comparing the result with a stored threshold. Updated decisions and syndromes return to the selected context; a completed frame leaves through the output register.

The WiMAX $(1056,528)$ implementation uses circulants of size $Z=44$. The $P=132=3Z$ lanes process three block columns at a time. A frame therefore contains $G=N/P=8$ groups. Four contexts, $C=4$, hold separate frame states and share the same processing lanes. Lane $\ell$ in group $g$ addresses node
\begin{equation}
 k=Pg+\ell,\qquad 0\leq g<G,\quad 0\leq\ell<P,
 \label{eq:lane_mapping}
\end{equation}
using zero-based hardware indices. Each lane-indexed memory contains $CG=32$ entries. Channel values, threshold indices, and smoothing counts use distributed memory, while decisions and syndromes have separate read and write buffers. Table~\ref{tab:archparams} summarizes the dimensions. Memory-bank division and structured QC processing provide related approaches to aligning node parallelism with memory access\cite{KimPark2013,Mhaske2015}.

\begin{table}[!htbp]
\centering
\caption{Architecture Dimensions}
\label{tab:archparams}
\small
\begin{tabular*}{\columnwidth}{@{\extracolsep{\fill}}lr@{}}
\toprule
Quantity & Value\\
\midrule
Base matrix / circulant size $Z$ & $12\times24$ / 44\\
VN lanes $P$ / groups per frame $G$ & 132 / 8\\
Frame contexts $C$ / state depth per lane $CG$ & 4 / 32\\
Input samples per accepted transfer & $22\times10$ bits\\
Sample transfers per column / group / frame & 2 / 6 / 48\\
Output width / transfers per frame & 44 bits / 24\\
Group acceptance-to-writeback latency & 5 clocks\\
Sustained group initiation interval & 1 clock\\
\bottomrule
\end{tabular*}
\end{table}
\subsubsection{Streaming initialization}
The initialization path in Fig.~\ref{fig:core} accepts 22 ten-bit channel samples per transfer. This width divides each 44-node block column into two halves and limits the width of the selectable alignment network. Six parallel 22-lane barrel shifters align the input with the check accumulators. The initial decisions, syndrome, and clipped check minima are formed by sign extraction, XOR, and minimum operations. A complete frame comprises 48 successful sample transfers, with admission and stalls governed by the interface handshake.

For a shift $p=22q+r$, the shifter rotates by $r=p\bmod22$; the half-column address accounts for $q$ and the carry across the 22-position boundary. Cyclic alignment is also used in QC decoder architectures\cite{KimPark2013,Mhaske2015}. The forward and gather-address relations are given in the supplement. Two setup buffers alternate between sample reception and weight generation, so reception of the next frame can overlap preparation of the current frame. A context is reserved at frame admission and becomes eligible for decoding after initialization.

After the final sample transfer, a population count of the initial syndrome gives $S_0$. Four ROMs return $w$, $\alpha$, $\lambda$, and $\eta$. The 528 check-node weights $W_i=w+\alpha\rho_i$ are computed using 22 DSP blocks, with the 44 minima of each block row processed in two halves. Each check node has one eight-word bank storing $+W_i$ and $-W_i$ for four contexts. During decoding, the syndrome selects the signed value and the QC connections distribute it to adjacent VN lanes. No weight multiplication is required in the iterative datapath. Earlier NGDBF hardware separates noise preprocessing from decoding at startup\cite{Sundararajan2016ASIC}; here, initialization prepares frame-dependent check weights.

\begin{figure*}[!t]
 \centering
 \includegraphics[width=0.94\textwidth]{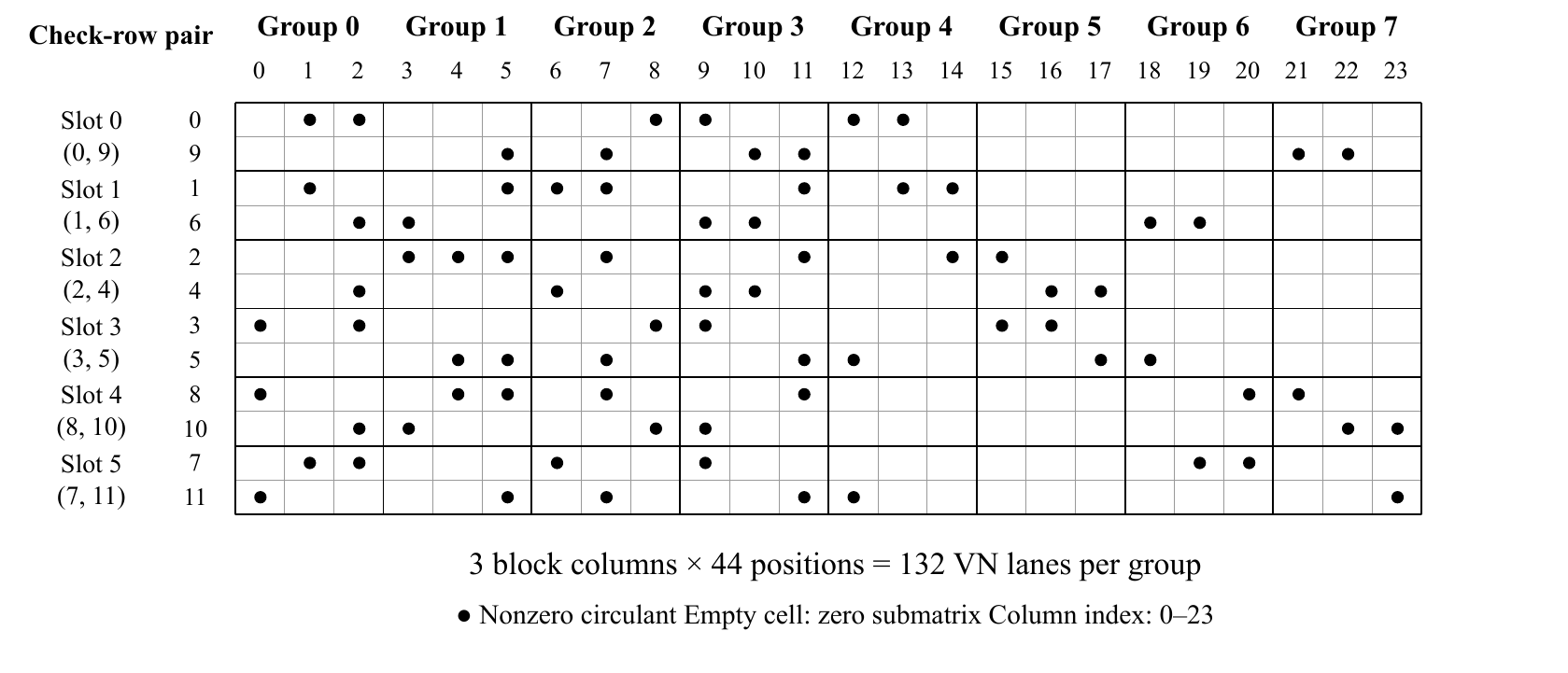}
 \caption{The selected WiMAX base matrix, with original block-row indices displayed in paired order. A dot denotes a nonzero $44\times44$ circulant; an empty cell denotes a zero submatrix. Each group contains three block columns, or 132 variable nodes. One block column is received in two 22-sample transfers.}
 \label{fig:qcmap}
\end{figure*}

\subsubsection{QC mapping}
The selected WiMAX base matrix $H_b$ has $12\times24$ entries. An entry of $-1$ represents an all-zero submatrix, whereas a nonnegative entry specifies a cyclically shifted identity matrix; in particular, zero specifies the identity. With variable position $c$ and shift $p$, the corresponding check position is $(c+p)\bmod44$. Figure~\ref{fig:qcmap} shows the nonzero circulants and the grouping used by the datapath.

The fixed matrix further determines the check-contribution slots. Define the support of block row $r$ by
\begin{equation}
 B_r=\{c:H_b(r,c)\ne-1\}.
 \label{eq:row_support}
\end{equation}
The paired rows shown in Fig.~\ref{fig:qcmap} have disjoint supports: for each pair $(a,b)$, $B_a\cap B_b=\varnothing$. A VN is therefore connected to at most one check in each pair, allowing its neighbors to be assigned to six input slots. Unused slots receive zero. Figure~\ref{fig:qcmap} shows the resulting assignment. The circulant connections are fixed at elaboration, and an eight-way selector chooses the active group. QC regularity similarly reduces interconnect complexity in earlier architectures\cite{Le2018,Mhaske2015}; the row pairs used here follow from the selected WiMAX matrix.

\subsection{Variable-Node Datapath and Threshold Lookup}
\begin{figure*}[!t]
 \centering
 \includegraphics[width=0.98\textwidth]{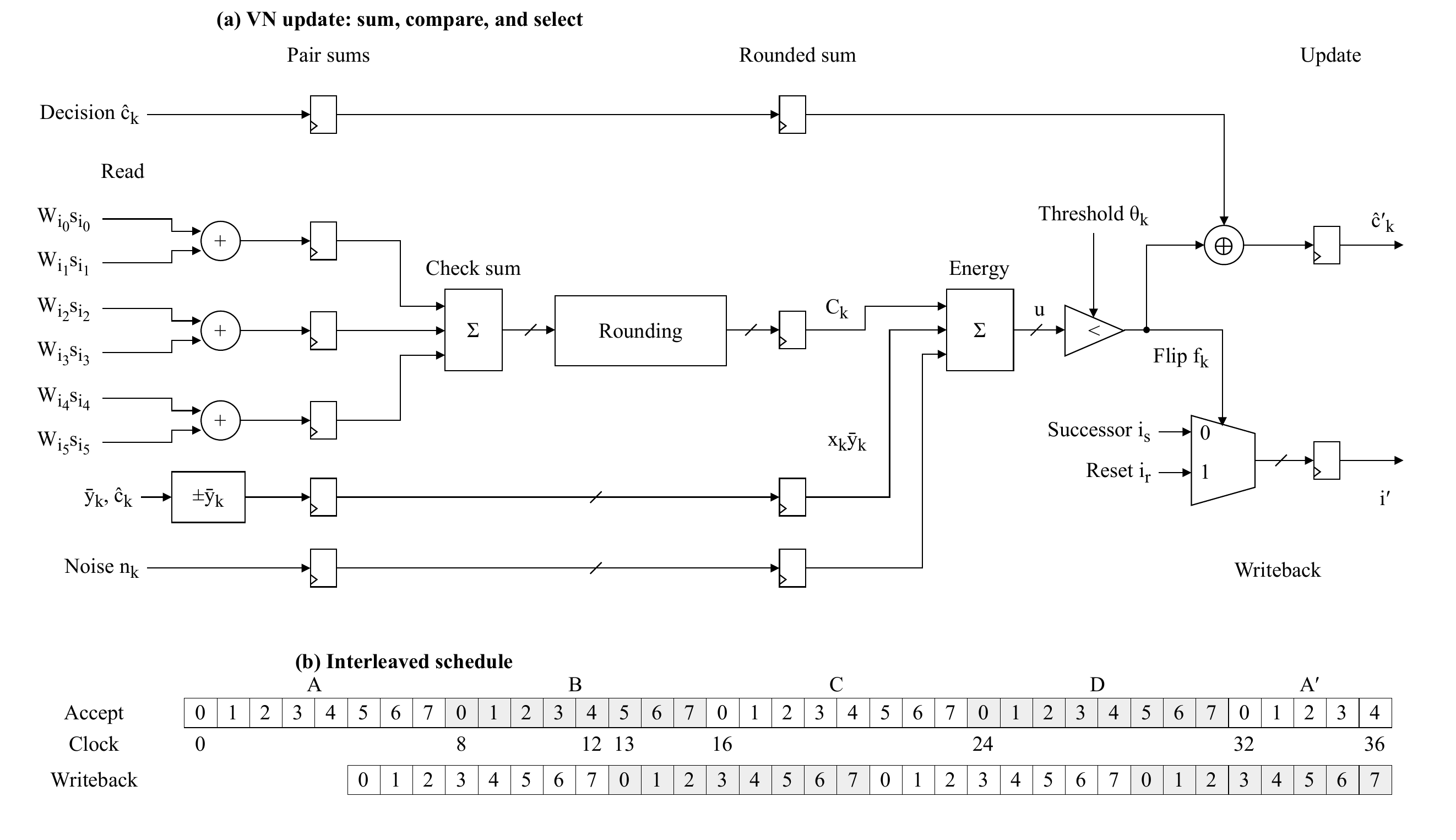}
 \caption{VN update and frame interleaving. (a) Inputs $W_{i_j}s_{i_j}$ correspond to neighboring checks $i_j\in\M(k)$; unused slots are zero. Rounding their sum from Q10 to Q7 gives $C_k$. The pre-saturation sum $u=C_k+x_k\bar y_k+n_k$ yields $f_k=\mathbf{1}[u<\theta_k]$. The XOR produces the next decision $\hat c_k^\prime$; the MUX selects the next threshold index $i^\prime$: reset $i_r$ if $f_k=1$, successor $i_s$ otherwise. Registers align operands; $\Sigma$ denotes addition. Iteration superscripts and Q-format subscripts are omitted. (b) Four ready contexts, without output or smoothing requests. Acceptance-to-writeback latency is five clocks.}
 \label{fig:pipeline}
\end{figure*}

The VN circuit in Fig.~\ref{fig:pipeline}(a) implements a sum, a threshold comparison, and two selections: whether to flip the decision and whether to advance or reset the threshold index. Six signed check contributions form three registered pair sums, followed by a total sum and one rounding operation. The channel term and perturbation are then added, and the comparison controls the decision XOR and index selection. State writeback completes five clocks after group acceptance; alignment registers keep all operands associated with the same node.

A fixed-point code $QF$ has $F$ fractional bits, and $\rnd$ rounds to the nearest integer with ties away from zero. Widths and numerical equivalence are detailed in Section~\ref{sec:fixedpoint}.

Threshold updates also use lookup in place of repeated multiplication. For a fixed frame coefficient, rounding restricts the decay trajectory to a finite set of values; a stored index identifies the current value and its successor. Quantized NGDBF threshold adaptation has been implemented using precomputed non-flip event tables\cite{Sundararajan2014}. Here, the rounded recurrence for each frame coefficient is encoded directly by its reachable states. Here, $\theta_I$ is the Q7 integer threshold code, and $\lambda_{Q10}$ is the decay coefficient encoded with ten fractional bits. The one-step map
\begin{equation}
 d_\lambda(\theta_I)=\rnd(\lambda_{Q10}\theta_I/1024)
 \label{eq:encoded_decay}
\end{equation}
generates trajectories from the initial value $-128$ and the post-flip value $-69$. Equal threshold values share one state, with a stored value and a successor index. A retained decision selects the successor; a flip selects the reset index. The transition table includes every rounded step and its fixed points.

\begin{figure*}[!t]
 \centering
 \includegraphics[width=0.92\textwidth]{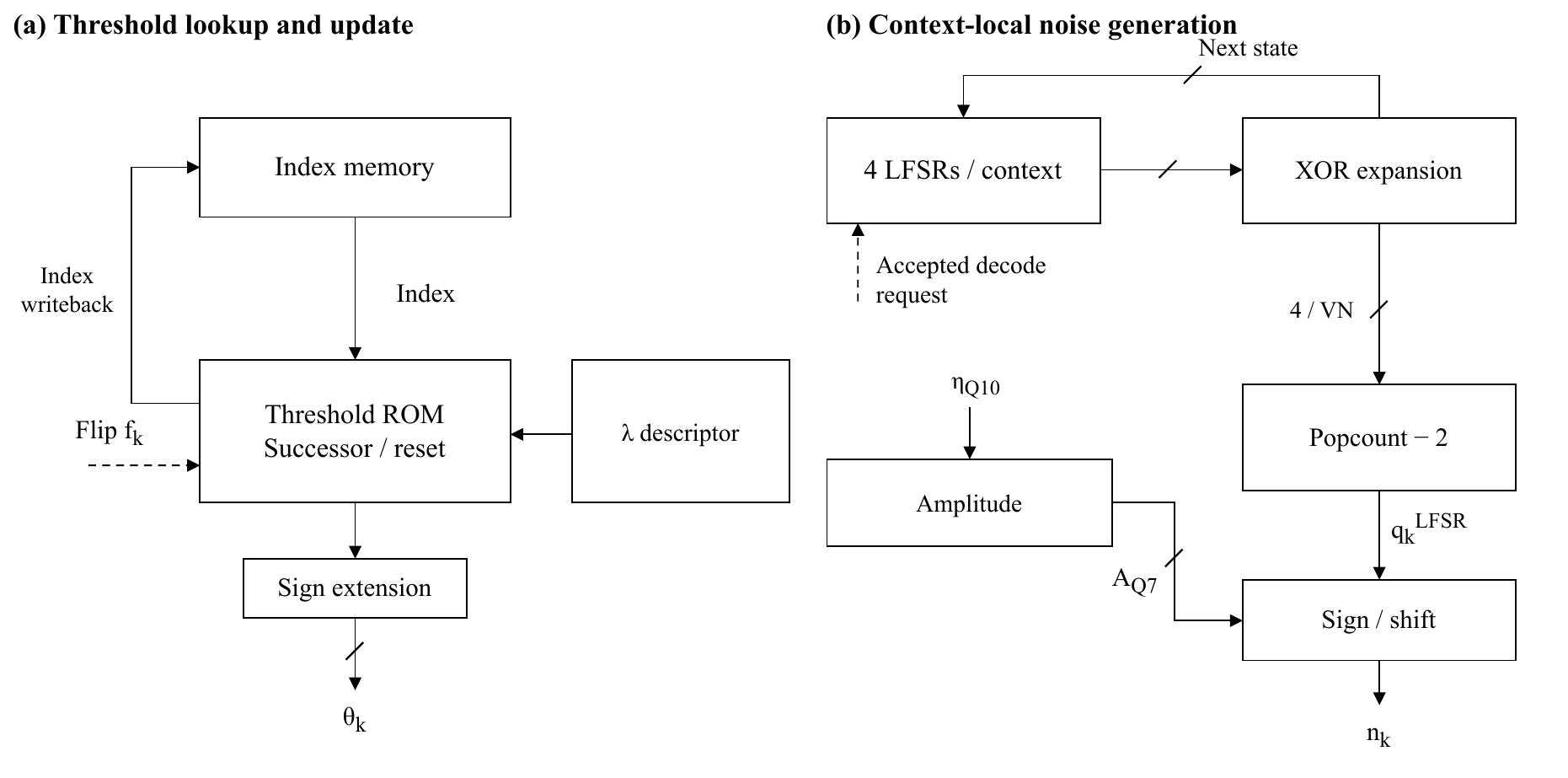}
 \caption{Threshold lookup and noise generation. (a) The stored index selects the current threshold. A flip selects the reset index for writeback; otherwise, the successor is selected. (b) Four LFSRs each provide one bit per VN. Popcount minus two produces $q_k^{\mathrm{LFSR}}\in\{-2,-1,0,1,2\}$; sign/shift selection scales it by the Q7 amplitude derived from $\eta$. Amplitude rounding uses ties away from zero. Only accepted decoding requests advance the generator states.}
 \label{fig:state_rng}
\end{figure*}

Figure~\ref{fig:state_rng}(a) shows the lookup circuit. The frame coefficient selects a descriptor, and each node stores a six-bit index into its threshold trajectory. At most 56 states are needed per coefficient. A $4096\times8$ value ROM is replicated 66 times: two read ports per copy supply the 132 lanes. Successor logic holds, increments, or redirects the index according to the trajectory; a flip selects its reset entry. The supplement gives the table construction.

\subsection{Parallel Perturbations and Synchronous State Updates}
\label{sec:perturbations}
Four 32-bit Galois linear-feedback shift registers (LFSRs) supply one bit each to form
\begin{equation}
 q_k^{\mathrm{LFSR}}=\sum_{r=0}^{3}b_{r,k}-2\in\{-2,-1,0,1,2\}.
 \label{eq:lfsr}
\end{equation}
Under the independent equiprobable-bit model, the probabilities are $(1,4,6,4,1)/16$, giving zero mean and unit variance. The scaled perturbation is
\begin{equation}
 A_{Q7}=\rnd(\eta_{Q10}/8),\qquad n_{k,Q7}=A_{Q7}q_k^{\mathrm{LFSR}}.
 \label{eq:noise}
\end{equation}
The implementation selects $0$, $\pm A_{Q7}$, or $\pm2A_{Q7}$ by sign and shift operations.

Each frame must consume the same perturbation sequence regardless of its scheduling position. Four 32-bit Galois LFSR states are therefore stored per context. Parallel expansion uses powers of the binary state-transition matrix, as in concurrent LFSR generation\cite{Milovanovic2015}. Let $A_r$ denote the one-step transition of generator $r$ over $\mathrm{GF}(2)$, $\bm\xi_{r,g}$ its state before group $g$, and $\bm e_0$ the least-significant-bit selector. The lane samples and next group state are
\begin{align}
 b_{r,g,\ell}&=\bm e_0^{\mathsf T}A_r^{\ell}\bm\xi_{r,g},\quad0\leq\ell<132,\label{eq:rng_sample}\\
 \bm\xi_{r,g+1}&=A_r^{132}\bm\xi_{r,g}.\label{eq:rng_advance}
\end{align}
The fixed binary maps are implemented as XOR networks. In each lane, a population count of four generator bits produces the five-level value in (\ref{eq:lfsr}). Sign and shift selection then gives $0$, $\pm A_{Q7}$, or $\pm2A_{Q7}$, with $A_{Q7}=\rnd(\eta_{Q10}/8)$. Figure~\ref{fig:state_rng}(b) shows the parallel expansion and amplitude selection.

An accepted decoding group stores the next state in (\ref{eq:rng_advance}) for the next group of the same context, including the transition from group 7 to group 0 of the next iteration. Output and smoothing requests leave this state unchanged. The reference model uses the same sample-before-step convention, nonzero initial states, and feedback masks. The supplement lists the initial states and feedback masks.

On the first iteration, the state-read path supplies the channel sign as the decision and threshold-state index 0, denoting the initial Q7 code $-128$. These substitutions avoid a separate initialization pass over the decision and threshold-index memories. Decisions and syndromes are double-buffered so that every group in an iteration reads the same committed state. Each group writes 132 updated decisions to the next buffer. With $H_g$ denoting the columns belonging to group $g$, the next syndrome is accumulated as
\begin{equation}
 \bm z^{(t)}=\bigoplus_{g=0}^{7}H_g\hat{\bm c}^{(t)}_g\pmod2.
 \label{eq:group_syndrome}
\end{equation}
A completion mask records their writebacks; after the final contribution, the buffer roles are exchanged and the completed syndrome is tested. Context and generation identifiers accompany requests so that writeback remains associated with the correct frame when a context is reused.

If $n_{0,k}$ counts zero decisions over the final 64 iterations, then
\begin{equation}
 v_k=n_{0,k}-(64-n_{0,k})=2n_{0,k}-64.
 \label{eq:vote_count}
\end{equation}
The output is determined by whether the zero count $n_{0,k}$ is below, equal to, or above 32. A six-bit counter saturated at 33 preserves this comparison while accumulating post-update decisions over iterations 237--300. Ties retain the last decision. The smoothing pass reuses the decision-writeback and syndrome paths.

\subsection{Frame-Interleaved Scheduling}
Groups within one iteration read the same committed state and can be accepted consecutively. The dependency occurs at the iteration boundary: after eight group requests, the final updates are still in the five-clock pipeline. The scheduler issues work from another ready context until all eight syndrome contributions of the first context have been committed. Multi-frame sharing is also used in QC-LDPC decoders\cite{Hasani2022} and dynamic layered schedules\cite{Kumawat2015}.

The scheduler issues groups 0--7 of an eligible context consecutively, then searches for another eligible context in round-robin order. In Fig.~\ref{fig:pipeline}(b), A0--A7 enter at clocks 0--7 and B0 follows at clock 8. A7 writes back at clock 12; its registered completion is visible to the scheduler at clock 13. C and D start at clocks 16 and 24. A can start its next iteration at clock 32, while D7 remains in flight until clock 36. The pipeline accepts one group per clock during sustained decoding.

An initially zero syndrome bypasses iteration; a zero syndrome after an update sends the context to output. Reaching iteration 300 with a nonzero syndrome selects smoothing. Completed decisions are read into a 132-bit holding register and sent as three 44-bit transfers per group, giving 24 transfers per frame. Output, smoothing, and decoding share the state-read path. Their arbitration, input readiness, and iteration dependencies are included in the elapsed batch time. A frame identifier associates each output with its input when convergence changes completion order.

\subsection{Fixed-Point Arithmetic}
\label{sec:fixedpoint}
The F7 representation selected in Section~\ref{sec:precision_selection} uses seven fractional bits for channel samples and the VN metric. Table~\ref{tab:formats} lists the stored formats and arithmetic widths.

\begin{table*}[t]
\centering
\caption{Datapath and State Representations}
\label{tab:formats}
\small
\renewcommand{\arraystretch}{1.08}
\begin{tabularx}{\textwidth}{@{}L{0.26\textwidth}ccY@{}}
\toprule
Quantity & Width (bits) & Fractional bits & Representation or operation\\
\midrule
Channel $\bar y_k$ & 10, signed & 7 & Per-node storage; clipping at $\pm2.5$\\
Check minimum $m_i$ & 8, unsigned & 7 & Initialization minimum; clipping at 1.5\\
$w,\alpha,\eta$ / $\lambda$ & 11 / 10, unsigned & 10 & Four $S_0$-addressed parameter ROMs\\
Signed weight $\pm W_i$ & 12, signed & 10 & Two signed entries per check node and context\\
Paired / complete check sum & 13 / 15, signed & 10 & Six-input adder tree\\
Scaled check sum $C_k$ & 12, signed & 7 & One rounding after total accumulation\\
Perturbation $n_k$ & 16, signed & 7 & Five-level sign and amplitude selection\\
Threshold value / state index & 8, signed / 6, unsigned & 7 / 0 & ROM value / per-node state\\
Metric comparison path & 17, signed & 7 & Sum before reference saturation\\
Smoothing zero count & 6, unsigned & 0 & Saturation at 33\\
Decision / syndrome & 1 each & 0 & Two buffers per context\\
LFSR state & 32 each & 0 & Four generators per context\\
Iteration / group index & 9 / 3, unsigned & 0 & Iterations 0--300 / groups 0--7\\
\bottomrule
\end{tabularx}
\end{table*}

A subscript $QF$ denotes an integer code with $F$ fractional bits, representing the code multiplied by $2^{-F}$. The perturbation amplitude in the main scale is
\begin{equation}
 A_{QF}=\rnd\!\left(2^F\eta_{Q10}/2^{10}\right).
 \label{eq:amplitudeF}
\end{equation}
For WiMAX F3, the complete amplitude mapping produces integer code 5. F7 retains codes 77--88, allowing frame-to-frame differences in $\eta$ to affect the applied perturbations. Threshold recursion also depends on the step size. For an integer threshold $\theta_I$, the condition
\begin{equation}
 (1-\lambda)|\theta_I|<\tfrac12
 \label{eq:stall}
\end{equation}
can make one decay step round back to the same value. Such fixed points are discussed in the NGDBF quantization analysis\cite{Sundararajan2014}. For example, $\lambda_{Q10}=922$ and $\theta_I=-4$ give approximately $-3.60$ before rounding and return to $-4$. This code represents $-0.5$ in Q3 and approximately $-0.0313$ in Q7.

A signed two's-complement format $\Sfmt{W}{F}$ has $W$ total bits, including the sign bit, and $F$ fractional bits; $\Ufmt{W}{F}$ denotes the unsigned format. The reference parameter mapping is
\begin{equation}
\begin{split}
 g_{Q10}=\operatorname{clip}\big(&\rnd(a_{g,Q20}S_0/2^{10})+b_{g,Q10},\\
 &[g_{\min,Q10},g_{\max,Q10}]\big).
\end{split}
\label{eq:intmap}
\end{equation}
Its constants are quantized from the full-precision mapping: the slope has 20 fractional bits, and the intercept and clipping limits have ten. All 529 possible syndrome weights, $S_0=0,\ldots,528$, are evaluated in advance. Four parameter read-only memories (ROMs), addressed by $S_0$, implement (\ref{eq:intmap}) during frame initialization.

For the clipped minimum code $m_{i,Q7}$, the effective check-node weight is represented by
\begin{align}
 r_{i,Q10}&=\rnd\!\left(\alpha_{Q10}m_{i,Q7}/192\right),\label{eq:intreliability}\\
 W_{i,Q10}&=w_{Q10}+r_{i,Q10},\label{eq:intweight}
\end{align}
where $192=1.5\times2^7$. The adjacent signed contributions are accumulated in Q10 and rounded together:
\begin{equation}
 C_{k,Q7}=\rnd\!\left(\frac{\sum_{i\in\M(k)}W_{i,Q10}s_i}{8}\right).
 \label{eq:intcheck}
\end{equation}
The reference metric and threshold recurrence are
\begin{equation}
\begin{split}
 E_{k,Q7}=\sat_{[-32768,32767]}\big(&x_k\bar y_{k,Q7}\\
 &+C_{k,Q7}+n_{k,Q7}\big),
\end{split}
\label{eq:fixedmetric}
\end{equation}
\begin{equation}
 \theta_{k,Q7}^{(t)}=\sat_{[-512,511]}\!\left[
 \rnd\!\left(\frac{\lambda_{Q10}\theta_{k,Q7}^{(t-1)}}{1024}\right)\right]
 \label{eq:fixedthreshold}
\end{equation}
for a retained decision. Initial and post-flip threshold codes are $-128$ and $-69$. Products retain intermediate precision until conversion, and the complete check sum is rounded once in (\ref{eq:intcheck}). The stored threshold can be reduced to eight bits because its reachable values lie in $[-128,-3]$.

The Q7 check sum is added to the channel term and perturbation on a signed 17-bit path. All reachable thresholds lie in $[-128,-3]$, inside the 16-bit saturation interval. Consequently,
\begin{equation}
 \sat_{[-32768,32767]}(u)<\theta_I\quad\Longleftrightarrow\quad u<\theta_I.
 \label{eq:comparison_equivalence}
\end{equation}
The signed 17-bit sum can therefore be compared directly with the sign-extended threshold.

\subsection{FPGA Results and Comparison}
The decoder is implemented with Vivado 2019.1 on an XC7A200T-FBG484-2 at 60 MHz. Table~\ref{tab:fpga} reports the decoder and complete test system from the same routed implementation. The 66 threshold-ROM copies use 66 36-kbit BRAMs, and four parameter ROMs use four 18-kbit BRAMs. Initialization accounts for the core's 22 DSP blocks.

\begin{table}[!htbp]
\centering
\caption{Post-Route FPGA Resource Use}
\label{tab:fpga}
\small
\begin{tabular*}{\columnwidth}{@{\extracolsep{\fill}}lrr@{}}
\toprule
Resource & Decoder core & Test system\\
\midrule
LUTs & 66,585 & 67,207\\
Flip-flops & 38,583 & 39,463\\
BRAMs (36-kbit equivalent) & 68 & 125.5\\
DSP blocks & 22 & 23\\
\bottomrule
\end{tabular*}
\par\smallskip
\begin{minipage}{\columnwidth}\footnotesize
The core includes initialization, memories, processing, scheduling, and output. The test system also includes preloaded vectors, clocking, serial control, and comparison logic.
\end{minipage}
\end{table}

Across seven SNR points, the RTL matched the fixed-point reference on 128 quantized inputs per point in the decoded words and recorded status fields. Additional directed tests exercise output stalls, initially zero syndromes, and context reuse.

FPGA measurements use 128-frame batches at 2.50 and 4.00 dB. Preloaded samples are supplied according to decoder readiness, with a continuously ready output receiver.

For $B$ completed frames and $C_{\mathrm{batch}}$ elapsed cycles, coded throughput is
\begin{equation}
 T_{\mathrm{coded}}=\frac{BNf_{\mathrm{clk}}}{C_{\mathrm{batch}}}.
 \label{eq:batch_throughput}
\end{equation}
The measured interval includes sample loading, initialization, decoding, smoothing, output transfers, and the final receiver comparison. Host communication is outside this interval. Mean frame latency is measured between frame acceptance and final output, separately from the elapsed time of the overlapping batch.

\begin{figure}[!htbp]
 \centering
 \includegraphics[width=\columnwidth]{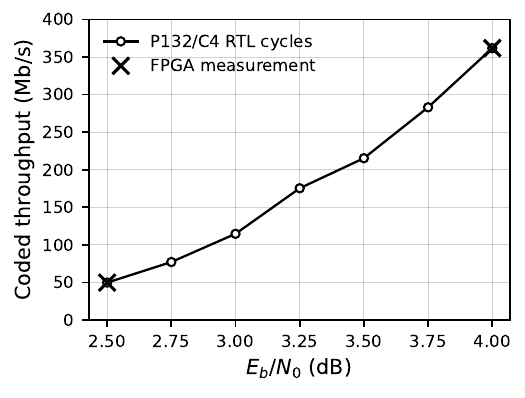}
 \caption{Coded throughput at 60 MHz. Open circles show the seven RTL batch results; crosses show FPGA measurements at 2.50 and 4.00 dB with the same inputs. Each batch contains 128 frames.}
 \label{fig:throughput}
\end{figure}

Throughput increases as fewer iterations are required (Fig.~\ref{fig:throughput} and Table~\ref{tab:throughput}), reaching 361.9 Mb/s at 4.00 dB compared with 49.7 Mb/s at 2.50 dB. Both values include initialization and output.

\begin{table}[!htbp]
\centering
\caption{128-Frame Batch Performance at 60 MHz}
\label{tab:throughput}
\footnotesize
\renewcommand{\arraystretch}{1.1}
\begin{tabular*}{\columnwidth}{@{\extracolsep{\fill}}rrrr@{}}
\toprule
$E_b/N_0$ & Mean & Throughput & Mean latency\\
(dB) & iterations & (Mb/s) & ($\mu$s)\\
\midrule
2.50$^\ast$ & 157.7 & 49.7 & 84.5\\
2.75 & 101.4 & 77.1 & 54.7\\
3.00 & 67.5 & 114.7 & 36.5\\
3.25 & 43.8 & 175.3 & 23.8\\
3.50 & 35.3 & 215.4 & 19.5\\
3.75 & 26.5 & 283.1 & 14.8\\
4.00$^\ast$ & 20.2 & 361.9 & 11.4\\
\bottomrule
\end{tabular*}
\par\smallskip
\begin{minipage}{\columnwidth}\footnotesize
$^\ast$FPGA-measured batch counts, also reproduced by RTL. Other throughput points use 132-lane RTL cycles at 60 MHz. Mean latency uses the RTL frame-acceptance and last-output events. The receiver is continuously ready; input transfers follow decoder readiness.
\end{minipage}
\end{table}

Table~\ref{tab:related_fpga} places the implementation alongside AWGN MNGDBF\cite{Li2025}. MNGDBF forms perturbations from transformed and cyclically shifted channel samples. HARS instead generates context-local perturbations and precomputes both signed check weights and threshold trajectories, concentrating iterative processing in selection, addition, and comparison. The table specifies the code, precision, and timing basis of each implementation: MNGDBF reports throughput for 100 iterations, while HARS uses measured batches with early termination and initialization and output included.

\begin{table*}[t]
\centering
\caption{FPGA Bit-Flipping Implementations and Reported Operating Conditions}
\label{tab:related_fpga}
\small\renewcommand{\arraystretch}{1.0}
\begin{tabularx}{\textwidth}{@{}L{0.24\textwidth}YY@{}}
\toprule
Item & MNGDBF, Li \emph{et al.} (2025)\cite{Li2025} & HARS (this work)\\
\midrule
Target FPGA & Not specified & XC7A200T-FBG484-2\\
Code / channel & PEGReg $(1008,504)$ / AWGN & WiMAX $(1056,528)$ / AWGN\\
Quantization & 6 bits total & 10-bit channel, 7 fractional bits\\
Clock frequency & 200 MHz & 60 MHz\\
Iteration schedule & 10 clocks per iteration & 8 group requests; 5-clock pipeline; 4 contexts\\
LUTs / flip-flops & 78,981 / 46,215 & 66,585 / 38,583\\
BRAM36-equivalents / DSPs & Not reported & 68 / 22\\
Perturbation source & Transformed received samples, cyclic shift & Four LFSRs per context, 132-sample expansion\\
Coded throughput (Mb/s) & 201.6 & 49.7 / 361.9\\
Throughput basis & Calculated with 100 iterations & FPGA, 128 frames at 2.50 / 4.00 dB\\
Mean iterations in measured batch & --- & 157.7 / 20.2\\
Timing interval & 1000 core clocks (100 iterations) & Input loading through final receiver comparison\\
\bottomrule
\end{tabularx}
\end{table*}

Other BF architectures emphasize different processing structures. Information-storage bit flipping (ISBF) stores prior energy values to shorten its update path and reports 90-nm CMOS synthesis\cite{Cui2020}; the variable-node-shift architecture (VNSA) distributes QC processing among heterogeneous units and reports ASIC synthesis for a binary symmetric channel\cite{Le2018}. Code, channel, precision, and throughput interval are the principal comparison dimensions identified in FPGA LDPC surveys\cite{Hailes2016}.

\section{Conclusion}
HARS combines check-level channel reliability with frame-dependent parameter selection in a synchronous bit-flipping decoder. Local weighting distinguishes parity constraints that would otherwise contribute equally, while the initial syndrome selects the operating parameters before iteration begins. The evaluated PEGReg and WiMAX configurations reduce BER and mean iteration count relative to the calibrated SM-NGDBF baseline. In the PEGReg ablation without retuning, local weighting provides the larger BER improvement, with a further reduction from frame adaptation.

The same separation between initialization and iteration guides the hardware design. Stored check weights and encoded threshold trajectories remove repeated multiplication from the VN update, and context-local perturbations preserve the sample sequence under frame interleaving. Four contexts sustain the shared 132-lane pipeline while retaining synchronous state updates. The resulting XC7A200T implementation reaches 361.9 Mb/s at 4.00 dB and 60 MHz, including initialization and output, demonstrating a direct hardware realization of the reliability-weighted update.

\FloatBarrier
\bibliographystyle{IEEEtran}
\bibliography{references}
\end{document}